\documentclass{ws-procs9x6}
\newcommand{\be}{\begin{equation}}
\newcommand{\ee}{\end{equation}}
\newcommand{\ba}{\begin{eqnarray}}
\newcommand{\ea}{\end{eqnarray}}
\newcommand{\bc}{\begin{center}}
\newcommand{\ec}{\end{center}}

\begin{document}
\begin{flushright}
SI-HEP-2009-15 
\end{flushright}

\title{ HADRONIC  FORM FACTORS: COMBINING QCD 
CALCULATIONS  WITH ANALYTICITY\footnote{ Talk at the Workshop 
"Shifmania, Crossing the boundaries: Gauge dynamics at strong coupling", May 14-17,2009, Minneapolis, USA  } }

\author{A. KHODJAMIRIAN}

\address{Theoretische Physik 1, Fachbereich Physik,
Universit\"at Siegen,\\ D-57068 Siegen, Germany\\ 
E-mail:
khodjam@hep.physik.uni-siegen.de}



\begin{abstract}
I discuss  recent applications of QCD 
light-cone sum rules to various form factors
of pseudoscalar mesons. In this approach 
both soft and hard contributions to the form factors 
are taken into account. 
Combining QCD calculation with the
analyticity of the form factors, one enlarges the 
region of accessible  momentum transfers.

\end{abstract}


\bodymatter

\section{Introduction}
At the beginning of  this talk let me  
quote  Misha Shifman:\\
"Unlike some models whose relation to Nature 
is still a big question mark, Quantum Chromodynamics will stay with us forever." \cite{Shifman1}

In the strong coupling domain, the 
quark-gluon gauge dynamics of QCD manifests itself 
in a form of hadrons. A  comprehensive analytic description of 
hadrons and their interactions in  QCD remains  
an unsolved  problem that may  also stay with us forever. 
Today we have at our disposal only approximate methods and 
effective theories of hadrons, with 
an impressive progress achieved in the numerical simulation of 
QCD on the lattice. 

An approximate analytical calculation of hadronic 
observables  has been made possible  with 
the advent  of QCD  sum rules \cite{SVZ}. Not only 
the original SVZ method is  still being extensively used, but also its  
"offspring", the light-cone sum rules (LCSR)
\cite{lcsr}.  In what follows, I overview  applications
of LCSR to hadronic form factors. I will also discuss
a possibility  to enlarge the region of accessible momentum
transfers by employing the analyticity of the form factors.  

\section{Hadronic form factors}

The simplest hadronic form factors parameterize 
electroweak  transitions between two 
ground-state pseudoscalar mesons. 
A well-known example  
is the pion form factor  generated by the quark electromagnetic 
current:
\be
 \langle \pi(p+q)|j_\mu^{em}|\pi(p)\rangle=
(2p+q)_\mu F_\pi(Q^2)\,,
\label{eq:fpi}
\ee
where  $j_\mu^{em}=\frac23\bar{u}(x)\gamma_\mu u (x) - \frac13\bar{d}(x)\gamma_\mu d (x)$
and  $Q^2=-q^2$. 
This form factor remains a popular test ground of QCD models,
and is being used nowadays to probe the 
AdS/QCD approach (see, e.g.,\cite{Rad,Erlich}). 

An important constraint on any model of the pion form factor is  
the $Q^2\to \infty $ asymptotics \,\cite{Fpiasympt,LB}\,:
\be
 F_\pi(Q^2)|_{Q^2\to \infty} = \frac{8\pi\alpha_s f_\pi^2}{9Q^2}
  \Bigg|\int\limits_0^1\!\! du \frac{\varphi_{\pi}(u,\mu)}{1-u}\Bigg|^2\,,
\label{eq:fpias}
\ee
determined by the perturbative gluon exchange between the 
quark and antiquark constituents of the pion.  In this 
factorization formula, the long-distance dynamics below the 
separation scale $\mu\sim \sqrt{Q^2}$   is encoded in the  
the pion distribution amplitude (DA) $\varphi_\pi(u,\mu)$. 
The latter is defined as the twist-2 part of the quark-antiquark
vacuum-pion matrix element expanded near the light-cone:
\be
\langle\pi(p)|\bar{u}(x)[x,0]\gamma_\mu\gamma_5 d(0)|0\rangle_{x^2=0}=
-ip_\mu f_\pi\int_0^1du\,e^{iup\cdot x} \varphi_\pi (u,\mu)\,,
\label{eq:phipi}
\ee
where $[x,0]$ is the gauge factor and $\mu\sim 1/\sqrt{|x^2|}$. The pion DA is a universal object,
in a sense that  it also enters factorization 
formulae for  other pion form factors. The knowledge
of $\varphi_\pi(u,\mu)$  
is however not sufficient for a form factor calculation 
at finite $Q^2$. A major question remains to be addressed: 
how large are the soft contributions to the form factor (\ref{eq:fpi}) which start from $O(1/Q^4)$.

Switching from the electromagnetic to the weak flavour-changing current
$\bar{u}\gamma_\mu b$,  one gets the   
$B\to\pi$ transition form factor:
\begin{equation}
 \langle \pi(p)|\bar{u}\gamma_\mu b |B(p+q)\rangle=
(2p+q)_\mu f^+_{B\pi}(q^2)+..\,,
\end{equation}
where ellipses indicate the  
presence of the second form factor, due to nonconservation of the current.
The $c \to d,s$ weak transitions generate similar $D\to \pi,K$ form factors.   
The role of large scale $Q^2$ is taken now by the heavy quark mass  $m_Q$ ($Q=b,c$).
In these heavy-light form factors there is again an interplay of hard and 
soft quark-gluon interactions. One can write an  
asymptotic factorization formula, similar to (\ref{eq:fpias}),
containing $\varphi_\pi(u,\mu)$ together with the heavy-meson 
DA. Again, such an ansatz  
is  phenomenologically incomplete, because soft 
contributions have to be included. 
They play an even more important role in 
heavy-to-light  form factors, since they are not power suppressed 
(in $1/m_Q$) with respect to  the hard  factorizable part.

Heavy-light form factors are not only interesting 
hadronic objects. Their knowledge 
is of primary importance for flavour physics. For example,
the $B\to \pi$ form factor is  needed to extract 
the quark mixing parameter $|V_{ub}|$ from 
data on semileptonic decay $B\to \pi l \nu_l$, whereas   
$D\to \pi$ and $D\to K$ form factors 
are used to determine $|V_{cd}|$ and, respectively, $|V_{cs}|$. 
The exploration of current and future  
experimental data on exclusive 
$B$ and $D$ decays demands accurate theoretical predictions for
these form factors.  

\section{QCD light-cone sum rules }

The method of LCSR was described in detail in many papers,
(see, e.g. \cite{BBKR,BpiNLO,BZ04,DKMMO,KKMO} and the review \cite{CK}), hence I only give a very short outline.
The central object is the correlation function of two quark currents
between the vacuum and on-shell pion state.
To give a familiar example,  
the calculation of the $B\to\pi$ form factors
starts from the following expression:
\ba
\int\!\! d^4x e^{iqx}\!
\langle \pi(p)|T\{\overline{u}(x)\gamma_\mu b(x), \!
\overline{b}(0)im_b\gamma_5d(0)\}|0\rangle
\!= \!F(q^2\!,\!(p+q)^2)p_\mu \!+\!...\,,
\label{eq:corr}
\ea
where only the invariant amplitude multiplying 
$p_\mu$ is relevant for the form factor $f^+_{B\pi}(q^2)$.
At $q^2, (p+q)^2 \ll m_b^2$,
the $b$-quark propagating in the correlation function 
has a large virtuality and the product of $b$-quark 
fields is expanded near the light-cone $x^2\sim 0$,
leading to OPE for the correlation function, schematically:
\ba
\hspace{-0.3cm}F_{OPE}(q^2,(p+q)^2)= 
\sum\limits_{\Gamma_a}  \int d^4x e^{iqx}C_{(\Gamma_a)}(x^2,m_b)
\langle \pi(p)|\bar{u}(x)\Gamma_a d(0)|0\rangle\, &
\nonumber\\
+\sum\limits_{\Gamma_b}  \int d^4x e^{iqx}\!\int\limits _0^1 dv\,\tilde{C}^{\alpha\beta}_{(\Gamma_b)}(x^2, v,m_b)
\langle \pi(p)|\bar{u}(x)G_{\alpha\beta}(vx)\Gamma_b 
d(0)|0\rangle\,,
\label{eq:ope}
\ea
where gauge factors are not shown, $\Gamma_{a,b}$ are various combinations of Dirac 
matrices. In the above, the short-distance coefficients $C_{(\Gamma_a)}$ and $\tilde{C}_{(\Gamma_b)}$  
stemming from the $b$-quark propagator 
are calculated perturbatively, whereas
the vacuum-pion matrix elements
are expressed via pion light-cone DA's.
The term with $\Gamma_a=\gamma_\mu\gamma_5$ 
yields the twist-2 DA defined in (\ref{eq:phipi}) and, in addition,  the twist-4 DA's.
The terms with  $\Gamma_a=i\gamma_5,\sigma_{\mu\nu}\gamma_5$
yield twist-3 DA's. Furthermore, the soft gluon emitted from
the $b$-quark propagator enters quark-antiquark-gluon 
operators, shown in the second line in (\ref{eq:ope}). These
operators are decomposed in pion three-particle 
DA's of twist-3 and 4.
The dominant twist-2 and twist-3 contributions 
are calculated to $O(\alpha_s)$,  obeying collinear 
factorization \,\cite{BpiNLO,BZ04,DKMMO}\,. Subdominant 
twist-4 and three-particle contributions
are power suppressed,  justifying the  truncated
twist/Fock-state expansion. It is a future task to assess
also the twist-5 and 6 terms of this expansion.

Matching OPE and hadronic dispersion relation 
for the correlation function at $|(p+q)^2|\sim m_b\chi$,
where $\chi$ is an intermediate scale $\gg \Lambda_{QCD}$,  one
uses quark-hadron duality for the 
hadronic states above $B$ meson, obtaining: 
\ba
F_{OPE}(q^2,(p+q)^2)= 
\frac{2m_B^2f_B f_{B\pi}^+(q^2)}{m_B^2-(p+q)^2}+
\int\limits_{s_0^B}^{\infty}ds \frac{[\mbox{Im}F(s,q^2)]_{OPE}}{s-(p+q)^2}\,.
\label{eq:sr}
\ea
Substituting  in (\ref{eq:sr})  the amplitude 
$F_{OPE}$ calculated  from (\ref{eq:ope}), 
one obtains an approximate analytical answer 
for the form factor in terms of pion DA's convoluted with 
calculable coefficients. Inputs in this relation include 
the $b$ quark mass (in the $\overline{MS}$ scheme),  
$\alpha_s$, and the set of low-twist 
universal pion DA's. The decay constant
$f_B$  is determined from two-point QCD (SVZ) sum rule. 
For other details, see, e.g. \,\cite{DKMMO}\,.

Importantly, in the resulting LCSR for the form factor
$f^+_{B\pi}$,
the leading twist-2 and 3 terms start from $\alpha_s^0$,  
with NLO corrections being $O(\alpha_s)$ suppressed.
Thus, in LCSR both  soft (nonfactorizable) and 
hard (factorizable) contributions are taken into account and the soft one 
dominates. Importantly, higher twist contributions 
are suppressed by  $(\Lambda_{QCD}/m_b)$ and/or $(\Lambda_{QCD}/\chi)$.  Since the calculation is done at finite $m_b$, a transition from $b$ to $c$ quark
in (\ref{eq:corr}) is straightforward, yielding 
LCSR for $D\to \pi,K$ form factors \cite{BBKR,KRWWY,BallDpi,KKMO}. 

The universality of the method
goes even further. Forming a vacuum$\to$pion 
correlation function of  the currents 
$j_\mu^{em}$ and $\bar{u}\gamma_\rho\gamma_5 d$,  
one obtains LCSR  for the pion form factor  
$F_\pi(Q^2)$ \, \cite{BH,BKM,BijnAK}\,, 
valid at  $Q^2\gg \Lambda_{QCD}^2$. This sum rule predicts a  
substantial soft contribution at intermediate $Q^2$
and, simultaneously, reproduces the QCD asymptotics (\ref{eq:fpias}).  Furthermore, employing
vacuum$\to$ baryon correlation functions with a 
baryonic interpolating current and DA's, one is able to calculate various baryon form factors, e.g.,  the nucleon form factors\,\cite{nucleon}\,. 

The method of LCSR has certain limitations. First of all, 
there is no "direct access" to the hadronic form factor.
The analytical expression in a form of factorized 
OPE is obtained for the correlation function, 
whereas the form factor enters the pole term of the 
dispersion relation. One isolates 
it from the contributions 
of higher states to the dispersion relation, 
estimated using quark-hadron duality.
This approximation introduces a sort of  ``systematic'' uncertainty 
of the method. To keep it under control, one fits the threshold parameter $s_0^B$  in (\ref{eq:sr}) by calculating 
the $B$-meson mass from the same LCSR. 
A typical accuracy of the form factors, calculated by this method 
is estimated at the level of $\sim \pm 15\%$, by varying all input parameters and scales within their adopted interval. 
One of the main uncertainties is  the shape  of the twist-2 DA $\varphi_{\pi}(u)$  expressed via Gegenbauer moments.

The region of momentum transfer $q^2$ accessible for LCSR 
is restricted: $f_{B\pi}^+(q^2)$ is calculated at
 $q^2\!\ll\!(m_B-m_\pi)^2$,  practically at   
$q^2<12-14 ~\mbox{GeV}^2$,  
 $f_{D\pi}^+(q^2)$ at $q^2\simeq 0$, and $F_\pi(Q^2)$ 
at $Q^2\geq $1 GeV$^2$.
It is therefore an important task to access other regions of $q^2$
where  more data are available.
This is possible, provided one uses the analyticity of the form factors.

\section{Employing the analyticity}

Hadronic form factors are analytic functions of the momentum-transfer variable $q^2$. A typical dispersion relation
which follows from the analyticity of the $B\to \pi$ form factor:
\vspace{-0.2cm}
\begin{equation}
f^+_{B\pi}(q^2)=\frac{m_{B^*}f_{B^*}g_{B^*B\pi}}{2(m_{B^*}^2-q^2)}+
\frac1\pi\!\!\!\!\!\! \int\limits_{(m_B+m_\pi)^2}^\infty 
\!\!\!\! ds\frac{\mbox{Im} f^+_{B\pi}(s)}{s-q^2}\,,
\label{eq:disp}
\end{equation}
takes into account the singularities 
located on the positive real $q^2$ axis, the lowest one being
the ground-state $B^*$ pole (with the $B^*B\pi$ coupling 
defined as in  \,\cite{BBKR}). Starting from the threshold at 
$q^2=(m_B+m_\pi)^2$,
there  are branch points and poles, generated by 
hadronic continuum states and excited resonances 
with  $B^*$ quantum numbers. Note that,
due to the QCD asymptotics $f^+_{B\pi}(q^2\to \infty) \sim 1/q^2$ (similar to (\ref{eq:fpias})), 
there are no subtractions in (\ref{eq:disp}).  
Importantly, this dispersion relation 
is valid at any $q^2$. Hence, a practical way 
to enlarge  the accessible $q^2$-region is to match 
the LCSR  result to  (\ref{eq:disp})  
at $q^2\ll m_B^2$ and analytically continue
the dispersion relation. This however can only be done 
if a model/ansatz is introduced for the integral 
over the spectral density of higher states in (\ref{eq:disp}), 
e.g., an effective pole\, \cite{BK}\,.

A less model-dependent approach employs 
conformal mapping (for earlier uses see \,\cite{confmap}). 
One maps the complex $q^2$-plane where (\ref{eq:disp})
is valid, onto  the unit circle  $|z|<1$  in the plane of the new variable:
 $z(q^2,t_0)=\frac{\sqrt{t_+-q^2}-\sqrt{t_+-t_0}}{\sqrt{t_+-q^2}+
 \sqrt{t_+-t_0}} $,
 where $t_+=(m_B+m_\pi)^2$, and $t_0<t_+$ is a  parameter. 
There are  many  applications \cite{zparam} of 
this approach to $B\to \pi$ and other form factors, 
combined with perturbative QCD bounds
obtained from the unitarity for the 2-point correlation function.
In fact, the bounds are usually not restrictive, 
and, since the maximal values of $|z|$ corresponding to 
kinematical boundaries of  $B\to\pi$ or 
$D\to \pi,K$  transitions are rather small,
a simple Taylor series near $z=0$ suffices
to parameterize the form factor.
The last version of this series parameterization \cite{BCL}
advocates a truncated power expansion:
\vspace{-0.2cm}
\begin{equation}
f^+_{B\pi}(q^2)= \frac{1}{1-q^2/m_{B^*}^2}\sum\limits_{k=0}^{k_{max}} a_k\Big(z(q^2,t_0)\Big)^k\,,
\label{eq:zparam}
\end{equation}
(with certain constraints on  $a_k$), 
where the $B^*$ -pole near the threshold is isolated.
Three or four parameters are sufficient for a reasonably
accurate parameterization.
Employing (\ref{eq:zparam}), it is possible 
to go beyond the  region where a LCSR calculation 
is valid. For that one has to  fit the  coefficients $a_k$ to 
the LCSR result for the form factor 
in  the "trusted "region of  $q^2$. After that, transforming $q^2\to z$, 
one continues (\ref{eq:zparam}) over $z$ beyond  
the initial region and finally transforms the variable $z$ back to $q^2$.

\section{Recent results for heavy-light form factors}

\subsection{$B\to \pi$ form factor  and $|V_{ub}|$}

Let me first quote our update of the $B\to \pi$ form factor
calculated from LCSR \cite{DKMMO} with the result: 
$f^+_{B\pi}(0)=0.26^{+0.04}_{-0.03}$, where 
a recent very accurate determination
of   the $\overline{MS}$ $b$-quark mass \,\cite{mc} was used. 
In this calculation no attempt yet was done to 
use the analytical continuation. On the contrary, 
in order to diminish the theoretical error for the form factor
at $q^2=0$, the calculated form factor shape   
at $0< q^2<12 $ GeV $^2$ 
(the estimated region of validity of LCSR) was fitted 
to the $q^2$ -distribution in $B\to \pi l \nu_l$, 
measured by BABAR  collaboration \cite{Bpibabar}. This fit 
allowed to  tighten   the constraints on the 
Gegenbauer moments of the twist-2 pion DA , yielding :
\vspace{-0.1cm}
\be
\varphi _\pi(u,\mu)=6u(1-u)\Big(1+a_2(\mu)C_2^{3/2}(2u-1)+
a_4(\mu)C_4^{3/2}(2u-1)\Big)\,, 
\label{eq:phipifit}
\ee
with
$a_2(1\mbox{GeV})= 0.16\pm 0.01$,
$a_4(1\mbox{GeV})= 0.04\pm 0.01$ (neglecting $a_{6,...}$).
Note that these intervals are quite narrow and are within
broader  "world averages" (see e.g., \cite{BBL}).
Deviation of the DA from its asymptotic form is thus rather mild. 
Finally, in \cite{DKMMO}  the form factor at $q^2=0$ 
was used to extract $|V_{ub}|$ 
from the data on $B\to \pi l \bar{\nu}$.
This and  other  recent $|V_{ub}|$ determinations  from $B\to \pi l \nu_l$  are summarized in the following table:\\[2mm]
\begin{tabular}{|c|c|c|c|}
\hline
[Ref.] & $f^+_{B\pi}(q^2)$  & $f^+_{B\pi}(q^2)$  & 
 $|V_{ub}| \times 10^3$ \\ 
&calculation& input &\\
\hline
\cite{FNALMilc} & lattice  & - &  3.38$\pm$0.36\\
\hline
\cite{HPQCD}&  lattice  & - &  3.55$\pm$0.25$\pm0.50$ \\
\hline
\cite{BZ04}& LCSR & - & $3.5\pm 0.4\pm 0.1$ \\
\hline
\cite{FlynnN}&- & lattice $\oplus$ LCSR &  $3.47\pm 0.29\pm 0.03$\\
\hline
\cite{DKMMO}& LCSR  & -&   
$ 3.5\pm 0.4 \pm 0.2\pm 0.1$ \\
\hline
\cite{BCL} &- & lattice$\oplus$ LCSR &$3.54\pm 0.24$\\
\hline
\end{tabular}\\[2mm]
Importantly, both LCSR  and lattice QCD determinations  
are in a good agreement with 
$|V_{ub}|=
(3.5^{+0.15}_{-0.14})\times 10^{-3}$ inferred from the 
CKM unitarity triangle  fits \,\cite{CKMfit}\,.
Let me also mention a LCSR calculation \cite{DM} 
of $B\to K$ and $B_s\to K$ 
form factors (needed e.g., for models 
of rare exclusive $B$ decays).  
 
\subsection{ $D\to \pi , K$  form factors and $|V_{cd}|$,
$|V_{cs}|$}

Following the same calculational scheme as 
for $B\to \pi$~\cite{KKMO} and $B\to K$ ~\cite{DM}
transitions, using the same set of pion and kaon DA's
and $\overline{MS}$ c-quark mass from\, \cite{mc}\,, we calculated 
\cite{KKMO} the 
$D\to \pi$ and $D\to K$ form factors.
In the latter, the $SU(3)_{fl}$ violation is taken into account 
in $O(m_s)$. Predictions for these form factors from lattice 
QCD and LCSR  
are presented in this table:\\[2mm] 
 \begin{tabular}{|cc|c|c|}
\hline
Method &[Ref.] &$f^+_{D\pi}(0)$&$f^+_{DK}(0)$\\
\hline
Lattice QCD &\cite{APE01}&$0.57\pm 0.06\pm 0.02$& $0.66\pm 0.04\pm 0.01$ \\
&\cite{Aubin05}&$0.64\pm 0.03\pm 0.06$&$0.73\pm 0.03\pm 0.07$\\
&\cite{QCDSF09}&$0.74\pm 0.06\pm 0.04$&$0.78\pm 0.05\pm 0.04$\\
\hline
LCSR&\cite{KRWWY}&$0.65\pm 0.11$&$0.78^{+0.2}_{-0.15}$\\
&\cite{BallDpi}&$0.63\pm 0.11$&$0.75\pm 0.12$\\
&\cite{KKMO} &$0.67^{+0.10}_{-0.07}$&$0.75^{+0.11}_{-0.08}$\\
\hline
\end{tabular}\\[2mm]
Our calculation was then 
used to determine 
$|V_{cd}|$ and $|V_{cs}|$ with an improved accuracy
from the recent CLEO data \cite{CLEO2} on semileptonic $D$ decays.
(see \cite{KKMO} for more details and numerical results).

In addition, the series parameterization similar to 
(\ref{eq:zparam}) was used 
to enlarge the accessible $q^2$ region. 
The form factors were calculated at $q^2<0$ (still 
within the region  of validity of LCSR), and fitted  to the
$z$-parameterization. The  result was then analytically 
continued, to cover the whole kinematical region  
$0\leq q^2\leq (m_D-m_{\pi(K)})^2$ of $D\to\pi(K) l \nu_l$
decay. The predicted shape of $f^+_{D\pi}(q^2)$ normalized at $q^2=0$  
is compared with the one measured by CLEO ~\cite{CLEO2}
in Fig. 1.
\begin{figure}[t!]
\includegraphics[scale=0.23]{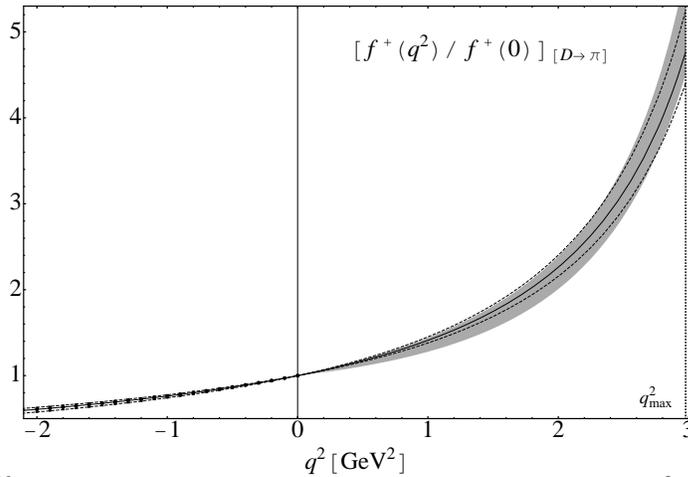}
\centering
\vspace{-0.3cm}
\caption{\it \cite{KKMO}
The shape of the $D\to \pi$ form factor obtained from LCSR
at $q^2\leq 0$, fitted to series parameterization (solid line and dashed lines indicating uncertainties), and  
compared with the shape measured by CLEO \cite{CLEO2}
at $0  \leq q^2 \leq (m_D-m_\pi)^2$ (shaded region)\,. 
}
\label{fig:ffslopevsCLEO}
\end{figure}

\section{ Pion form factor}
As already mentioned, the LCSR approach allows 
one to calculate any hadronic form factor, provided the 
momentum transfer is sufficiently below the hadronic threshold
in the channel of the transition current.  
It is therefore interesting to return to the pion e.m. form factor
$F_\pi(Q^2)$ which is quite sensitive to the pion twist-2 DA  and
recalculate it  from LCSR \,\cite{BKM,BijnAK}\,. In this sum rule
the  twist-2 term with $O(\alpha_s)$ corrections and twist-4,6 
terms are taken into account. With the pion DA  
given by (\ref{eq:phipifit}) and taking the remaining 
input from \,\cite{BijnAK}\,, I recalculated 
the pion e.m. form factor at $Q^2=1.0-5.0$ GeV$^2$.
The result is presented in Fig. 2, for the central 
values of the input. Conservatively, a $\pm 15\%$  uncertainty 
still has to be added. 
\begin{figure}[t!]
\centering
 \includegraphics[scale=1.2]{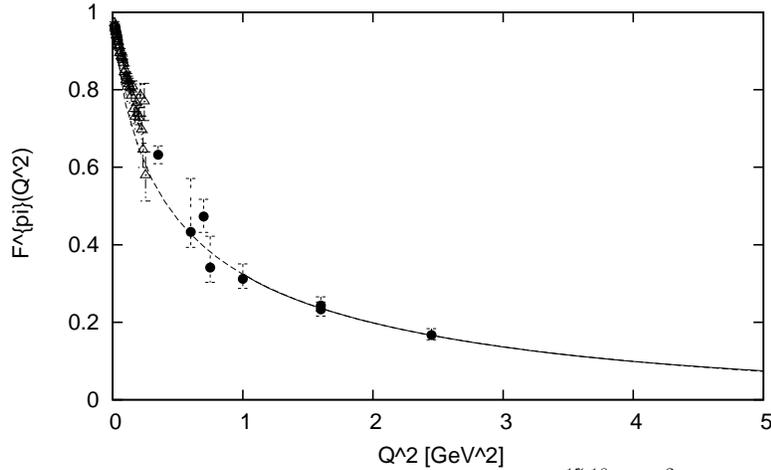}
\vspace{-0.3cm}
\caption{\it
The pion e.m. form factor calculated from LCSR \cite{BKM,BijnAK} 
at $Q^2=1-5 $ GeV$^2$ (solid), fitted to series parameterization
and analytically continued to smaller $Q^2$ (dashed), 
compared with JLAB data \cite{JLAB} (filled points)
and NA7 data\cite{Amendolia}(open triangles)}  .
\label{fig:ffpion}
\end{figure}
The agreement with the recent accurate data at $Q^2\leq 2.45$~GeV$^2$, 
obtained at Jefferson Lab \cite{JLAB} is encouraging.
     
Furthermore, it is easy to  write down an analog of series
parameterization (\ref{eq:zparam}) for $F_\pi(Q^2)$ , which is simply a polynomial in $z$ in this case
(the lowest $\rho$-meson pole is far above the threshold)
and continue the form factor to small $Q^2$. 
The result shown in Fig.~2 is consistent with 
the  direct  measurement of the pion form factor 
in this region \cite{Amendolia}
and correctly reproduces $lim_{Q^2\to 0}F_\pi(Q^2)= 1$ (within the accuracy of the sum rule). 

I conclude that the pion DA (\ref{eq:phipifit}) 
used in calculating the heavy-light form factors 
successfully passes the pion form factor test.  
A more detailed study  will be published elsewhere. 

\section{ Photon-pion transition form factor}
\vspace{-0.2cm}
{\it (a comment added after the talk)}\\

 The pion twist-2 DA also  determines the QCD asymptotics 
\cite{LB,BL2} of the  
$\gamma^* \gamma^*\to \pi^0$ amplitude:
\vspace{-0.3cm}
\ba 
F^{\gamma^*\pi}(Q^2,k^2)=\frac{\sqrt{2}f_\pi}{3}
\int\limits_0^1 du \frac{\varphi_\pi(u,\mu)}{Q^2(1-u) +|k^2|u}~\oplus ~
O(\alpha_s)\,,
\label{eq:pigamas}
\ea
where both photon virtualities, $Q^2=-q^2$ and 
$|k^2|\neq Q^2 $, are  sufficiently large. 
The $\gamma^* \gamma^*\to \pi^0$ amplitude 
is however only measured \cite{CELLO,CLEOgammapi,BABAR} 
when one of the photons is almost real ($k^2\simeq 0$), in which case
it reduces to the photon-pion transition form factor 
$ F^{\gamma\pi}(Q^2)=F^{\gamma^*\pi}(Q^2,0)$.
Note that at $k^2=0$ the factorization formula (\ref{eq:pigamas})
is incomplete,  because the real photon has
a long-distance, hadronlike  component.
Hence, substituting a certain 
model of $\varphi_\pi(u,\mu)$ in (\ref{eq:pigamas}) 
and comparing the result at $k^2=0$ with experimental data 
on $ F^{\gamma\pi}(Q^2)$, is not yet a conclusive test of QCD.
Calculations which take into account 
long-distance component of the photon can be 
found in the literature, (see, e.g. \cite{RR}).

A method to calculate $F^{\gamma\pi}(Q^2)$,
combining LCSR with the 
analytical continuation was suggested in \cite{AK99}\,.
The $\gamma^*\gamma^*\to \pi^0$ amplitude
was calculated using  light-cone OPE, 
with both  $Q^2=-q^2$ and $|k^2|$ large, adding 
also the twist-4 term to (\ref{eq:pigamas}).
(A calculation of the next, twist-6 term remains a future task.)
Equating this amplitude to the dispersion relation in $k^2$:
 \be
 F^{\gamma^*\pi}_{OPE}(Q^2,k^2)=
\frac{\sqrt{2}f_\rho F_{\rho\pi}(Q^2)}{m_\rho^2-k^2}+\frac{1}{\pi} \int_{s_0}^\infty 
\frac{\mbox{Im} F^{\gamma^*\pi}(Q^2,s)}{s-k^2}\,,
 \label{eq:fpig}
 \ee
 where $\rho$ and $\omega$ contributions are combined 
 in one resonance term with $m_\rho \simeq m_\omega$, 
 one calculates the $\rho\to \pi$ form factor using the technique
 of LCSR and employs quark-hadron duality for the integral
 above the threshold $s_0$.  After that a smooth transition
 to $k^2\to 0$ in the dispersion relation is possible, 
 yielding the form factor $F^{\gamma\pi}(Q^2)$. 
The $O(\alpha_s)$ corrections in this approach were calculated 
in\, \cite{SY}\,, and some recent updates  can be found in   
\,\cite{Ag,MS09}\,.
 
 Intrigued by the new BABAR data\,\cite{BABAR}\,, 
I returned to the calculation of   $F^{\gamma\pi}(Q^2)$
from (\ref{eq:fpig}), 
using the twist-2 pion DA   (\ref{eq:phipifit}),
the remaining  input from \,\cite{AK99}\,,
and adding the $O(\alpha_s)$ corrections
from \,\cite{SY}\,.
The result is shown in Fig.~3 (for the central values of 
the input). At large $Q^2$, as expected,   $Q^2F^{\gamma\pi}(Q^2)$ tends to a constant 
\footnote{Note that 
the $O(\alpha_s\ln^2Q^2)$ 
terms in the NLO part of the form factor\, \cite{SY}\, and 
the  effect of their resummation  still have to be assessed.}.
\begin{figure}[t!]
\centering
\includegraphics[scale=1.0]{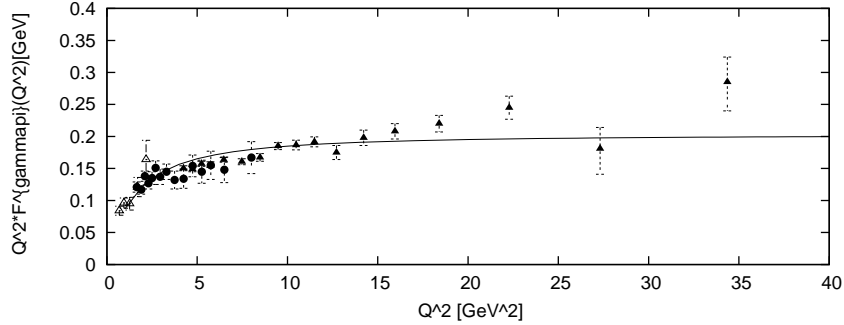}
\vspace{-0.5cm}
\caption{\it
Photon-pion form factor $Q^2 F^{\gamma\pi}(Q^2)$ calculated from 
a combination of LCSR and dispersion relation \cite{AK99} ,
including $O(\alpha_s)$ corrections\, \cite{SY}\,, compared 
with the CELLO\,\cite{CELLO} (open triangles), 
CLEO\,\cite{CLEOgammapi} (full points) and BABAR \,\cite{BABAR}
data (full triangles)}.
\label{fig:fpigam}
\end{figure}
In spite of a reasonable
 agreement with the old CELLO and CLEO data at low $Q^2$ 
 and with the new BABAR data up to $q^2\leq 15 $ GeV$^2$,
 this calculation (as well as the one in \cite{MS09}) does 
not reproduce the tendency of the increasing $Q^2F^{\gamma\pi}(Q^2)$ 
at larger $Q^2$ visible in the data, although the experimental 
errors in this region are rather large. 
An additional measurement, e.g.,  by Belle 
collaboration is important to confirm these very 
interesting data.

\vspace{-0.4cm}
\section*{Acknowledgments}
I am very grateful
to FTPI for inviting me to take part  
in the celebration of Misha Shifman's extraordinary
achievements in particle theory. Due to his important 
and enlightening contributions, deep connections 
between QCD and hadron phenomenology
were found and are being extensively used.\\
I acknowledge a useful discussion with Vladimir Braun.
This work is supported by the Deutsche Forschungsgemeinschaft
under the  contract No. KH205/1-2.

\end{document}